\documentclass[]{aastex631}

\usepackage{float}
\usepackage{amsmath}
\usepackage{bm}

\shorttitle{A Sandwich Model for CL-AGNs}
\shortauthors{Ma et al.}

\begin{document}
	\title{A Sandwich Model for Changing-Look AGNs}
	
	\author{Qian-Qi Ma}
	\author{Wei-Min Gu}
	\affiliation{Department of Astronomy, Xiamen University, Xiamen, Fujian 361005, People’s Republic of China; guwm@xmu.edu.cn}
	
	\author{Zhen-Yi Cai}
	\affiliation{CAS Key Laboratory for Research in Galaxies and Cosmology, Department of Astronomy, University of Science and Technology of China, Hefei 230026, People’s Republic of China}
	\affiliation{School of Astronomy and Space Science, University of Science and Technology of China, Hefei 230026, People’s Republic of China}
	
	\author{Xinwu Cao}
	\affiliation{Institute for Astronomy, School of Physics, Zhejiang University, 866 Yuhangtang Road, Hangzhou 310058, People’s Republic of China}
	
	\author{Jun-Xian Wang}
	\affiliation{CAS Key Laboratory for Research in Galaxies and Cosmology, Department of Astronomy, University of Science and Technology of China, Hefei 230026, People’s Republic of China}
	\affiliation{School of Astronomy and Space Science, University of Science and Technology of China, Hefei 230026, People’s Republic of China}
	
	\author{Zhi-Xiang Zhang}
	\author{Mouyuan Sun}
	\affiliation{Department of Astronomy, Xiamen University, Xiamen, Fujian 361005, People’s Republic of China; guwm@xmu.edu.cn}
	
	\begin{abstract}
		
		\noindent 
		The spectral variability of changing-look active galactic nuclei (CL-AGNs) occurred on timescales of years to tens of years, posing a significant challenge to the standard thin disk model. In this work, we propose a sandwich model, including an optically thick disk in the mid-plane (Disk 1) and two disks of low effective optical depth on both sides (Disk 2). These two types of disks are coupled with magnetic fields, which allow viscous torque interaction between them. As a consequence, the radial velocity of Disk 1 can increase by up to three orders of magnitude compared to the standard thin disk, leading to an equivalent decrease in the accretion timescale. Therefore, such a sandwich model can account for the rapid variability in CL-AGNs. In addition, we also discuss the influence of the magnetic pressure on Disk 2. When Disk 2 is dominated by the magnetic pressure, it resembles a ``warm corona", which is responsible for the soft X-ray excess.
		
	\end{abstract}
	
	
	\section{Introduction} \label{sec:intro}
	
	Active galactic nuclei (AGNs) are the bright cores of galaxies, and their radiation energy comes from the accretion of supermassive black holes (SMBHs). Based on the presence or absence of broad emission lines, AGNs are classified as Type 1 and Type 2 \citep{1943ApJ....97...28S}. According to the AGN unification model, differences in the width of emission lines are caused by the different sightline effect \citep[e.g.,][]{1995PASP..107..803U, 2006agna.book.....O}. According to this model, all AGNs should have clear types because the sightline remains unchanged. However, observations in the past few decades have shown that some AGNs experienced Seyfert-type transitions on the timescales of several months to several decades \citep[e.g.,][]{1986ApJ...311..135C, 2014ApJ...796..134D, 2016MNRAS.457..389M, 2018ApJ...862..109Y, 2019ApJ...883...94T}, and some even experienced multiple changes \citep[e.g.,][]{2016A&A...593L...8M, 2019MNRAS.483..558O, 2020ApJ...901....1W}. These AGNs are called changing-look AGNs (CL-AGNs). In addition to the variations of the broad emission lines, CL-AGNs often exhibit multi-wavelength variations. These variations were discovered in mid-infrared \citep[e.g.,][]{2017ApJ...846L...7S, 2018ApJ...864...27S, 2019ApJ...885..110Y, 2022ApJS..258...21W, 2021ApJS..252...32J}, UV \citep[e.g.,][]{2018A&A...618A..83Z, 2023MNRAS.521.3810N}, and X-ray \citep[e.g.,][]{2015ApJ...800..144L, 2018ApJ...866..123M, 2019MNRAS.483L..88P, 2022ApJ...930...46L, 2022JHEAp..33...20L} bands. 
	
	The physical mechanism of CL-AGNs remains a mystery. Several hypotheses have been proposed and some of them successfully explain specific CL-AGNs or a class of them. A straightforward explanation is that the broad line region is obscured by moving clouds or tori in the line of sight of the observer \citep[e.g.,][]{1989ApJ...346L..21G, 2012ApJ...747L..33E}, but only a few CL-AGNs can be explained by this scenario \citep[e.g.,][]{2015ApJ...803...57I, 2018ApJ...862..109Y, 2019ApJ...887...15W}. Another popular scenario is the tidal disruption event (TDE) \citep{2015MNRAS.452...69M}, which has successfully explained some CL-AGNs \citep[e.g.,][]{2017MNRAS.465L.114W, 2020ApJ...898L...1R, 2021ApJ...920L..25S, 2022MNRAS.517L..71Z}. However, TDE is not applicable to some repetitive CL-AGNs and the quasi-periodic outbursts \citep[e.g.,][]{2018ApJ...861...51K, 2019MNRAS.483L..88P, 2020A&A...644L...5H, 2022ApJ...936...75L}. In addition, close binaries of supermassive black holes (CB-SMBHs) with high eccentricities may trigger CL events \citep[e.g.,][]{2018ApJ...861...51K, 2020A&A...643L...9W}, but this scenario is only capable of the periodic CL events. \cite{2018ApJ...861...51K} proposed that Mrk 1018 is an oscillating recoiled SMBH which has a highly eccentric orbit with a period of 29 years, and \cite{2020A&A...643L...9W} suggested that some CL-AGNs are close binary systems and the CL timescales are related to the orbital periods. On the other hand, the dramatic random fluctuation of the extreme ultraviolet emission is another plausible explanation for CL-AGNs \citep[e.g.,][]{2018ApJ...855..117C, 2020ApJ...892...63C, 2020ApJ...891..178S}.
	
	In addition, the rapid changing of the accretion can also create CL events, which is consistent with their multi-wavelength variations and polarization invariance \citep[e.g.,][]{2016MNRAS.457..389M, 2016MNRAS.461.1927P, 2016A&A...593L...9H, 2018A&A...618A..83Z, 2018MNRAS.480.3898N, 2019A&A...625A..54H, 2022ApJ...930...46L, 2022ApJ...926..184J}. However, the accretion timescale for the standard thin disk \citep[SSD;][]{1973A&A....24..337S} is longer than $ 10^{3} $ yr in the UV/optical area, which is inconsistent with the observed CL timescale. Some other accretion models were proposed to explain the short timescales of CL events, which can be divided into stable and unstable models. For the former scenario, \cite{2019MNRAS.483L..17D} proposed that magnetically elevated disk has shorter inflow timescale. Following this hypothesis, \cite{2021ApJ...916...61F}, \cite{2023ApJ...958..146W} and \cite{2023MNRAS.526.2331C} suggested that the magnetic disk-outflows can effectively carry away the angular momentum and decrease the accretion timescale. For the latter scenario, \cite{2020A&A...641A.167S} put forward an ADAF (advection-dominated accretion flow)-SSD model which has a narrow unstable radial zone between ADAF and SSD. They argued that the repeating outbursts in some CL-AGNs are related to the radiation instability in the unstable zone. Based on this model, \cite{2021ApJ...910...97P} proposed that large-scale magnetic fields can significantly shorten the period of outbursts, explaining the quasi-periodic eruptions.
	
\begin{figure*}
	\gridline{\fig{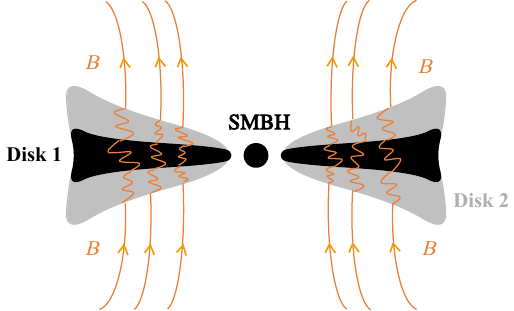}{0.45\textwidth}{(a)}
		\fig{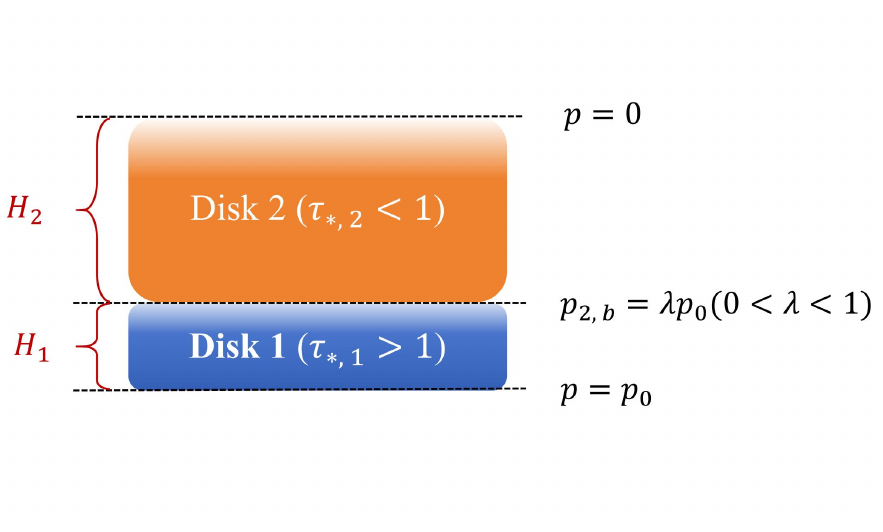}{0.45\textwidth}{(b)}
	}
	\caption{An illustration of the sandwich model. We propose that the Disk 1 and Disk 2 are coupled in the $ z $ direction via magnetic fields, as shown in (a), while (b) annotates several key parameters in the $ z $ direction. }
	\label{fig:1}
	\end{figure*}

	In this work, we construct a sandwich model, including an optically thick disk (Disk 1), with two optically thin disks (Disk 2) above it on both sides, as shown in Figure \ref{fig:1}. The ``optically thin'' and ``optically thick'' mentioned here refer to the effective optical depth, defined as $ \tau_{\ast} = \sqrt{\tau_{\mathrm{es}} \tau_{\mathrm{ff}}} $, where $ \tau_{\mathrm{es}} $ and $ \tau_{\mathrm{ff}} $ are electron scattering and free-free absorption optical depth, respectively. Disk 2 has higher radial velocity, which drags Disk 1 through magnetic coupling and increases the radial velocity of Disk 1. This mechanism can decrease the accretion timescale of Disk 1 by three to four orders of magnitude. In addition, we also discuss the impact of the magnetic pressure in Disk 2. When the magnetic pressure is negligible in Disk 2, its temperature will reach to $ 10^{9} $ K, which resembles a ``hot corona'' \citep[e.g.,][]{1993ApJ...413..507H, 2002ApJ...575..117L}. However, when the magnetic pressure is more than 10 times greater than the gas pressure, Disk 2 manifests itself a ``warm corona", which may explain the soft X-ray excess phenomenon in AGNs \citep[e.g.,][]{2015A&A...580A..77R, 2018MNRAS.480.1247K, 2022ApJ...937...31G, 2024MNRAS.530.1603B}. We describe our model and the basic equations in Section \ref{sec:2}. The numerical results and figures are presented in Section \ref{sec:3}. Our conclusions and discussion are presented in Section~\ref{sec:4}.

	\section{Model} \label{sec:2}
	
	We describe the model in the cylindrical coordinate system ($ r $, $ \phi $, $ z $), and the equations of Disk 1 and Disk 2 are presented in the following subsections. The symbols with subscript ``1'' and ``2'' respectively
	represent Disk 1 and Disk~2, and the subscript ``0'' represents the equatorial plane. The input and output parameters are listed in Tables \ref{tab:1} and \ref{tab:2}, respectively. 
	
	\begin{deluxetable*}{ccllc}[]
		\tablenum{1}
		\tablewidth{0pt}
		\tablecaption{Input parameters \label{tab:1}}
		\tablehead{
			\colhead{Parameter} & \colhead{Value} & \colhead{Meaning}
		}
		\startdata
		$ \alpha $ & 0.1 & viscosity parameter \\
		$ \beta_{\mathrm{m}} $ & $ 0 \sim 100 $ & the ratio of magnetic pressure in Disk 2 \\
		$ \eta $ & 0.1 & energy conversion efficiency \\
		$ \kappa_{\mathrm{es}} $ & $ \rm 0.4 \: cm^{2} g^{-1} $ & electron scattering opacity \\
		$ r $ & $ 50 \: r_{\mathrm{s}} $ & radius  \\
		$ M $ & $ 10^{7}, 10^{8} \: M_{\odot} $ & mass for SMBH \\
		$ \dot{M} $ & $ 0.01 \sim 0.1 \: \dot{M}_{\mathrm{Edd}} $ & accretion rate (total) \\
		$ \dot{M}_{2} / \dot{M} $ & $ 0.01 \sim 0.5 $ & accretion rate proportion of Disk 2 \\
		$ v_{1} / v_{2} $ & $ 10^{-3} \sim 1 $ & radial velocity ratio \\
		\enddata
		\tablecomments{The input parameters in our model. The symbols, values and meaning of the parameters are listed in the first, second, and third column, respectively.}
	\end{deluxetable*}
	
	\begin{deluxetable*}{cl}[]
		\tablenum{2}
		\tablewidth{0pt}
		\tablecaption{Output parameters \label{tab:2}}
		\tablehead{
			\colhead{Parameter} & \colhead{Meaning}
		}
		\startdata
		$ \Omega_{\mathrm{K}} $ & angular velocity \\
		$ \lambda $ & the ratio of the pressures \\
		$ I $  & interaction stress \\
		$ t_{\mathrm{acc, 1}} $ & accretion timescale of Disk 1  \\
		$ v_{1}, v_{2} $ & radial velocity  \\
		$ p_{0}, p_{\mathrm{2, b}} $ & pressure \\
		$ \rho_{0}, \rho_{\mathrm{2, b}} $ & density  \\
		$ T_{0}, T_{\mathrm{2, b}} $ & temperature  \\  
		$ H_{1}, H_{2} $ & half thickness \\
		$ \Sigma_{1}, \Sigma_{2} $ & surface density  \\
		$ \Pi_{1}, \Pi_{2} $ & pressure integration  \\
		$ T_{r\phi, 1}, T_{r\phi, 2} $ & vertically integrated stress  \\
		$ \tau_{\mathrm{es, 1}}, \tau_{\mathrm{es, 2}} $ & electron scattering optical depth \\
		$ \tau_{\mathrm{ff, 1}}, \tau_{\mathrm{ff, 2}} $ & free-free absorption optical depth \\
		$ \tau_{\ast, 1}, \tau_{\ast, 2} $ & effective optical depth
		\\
		\enddata
		\tablecomments{The Output parameters in our model. The symbols and meanings of the parameters are listed in the first and second column, respectively.}
	\end{deluxetable*}
	
	\subsection{Optically Thick Disk (Disk 1)}
	
	The equations of Disk 1 are based on the SSD model, and this disk can be described as a ``radially faster SSD". We assume this disk maintains Keplerian rotation and the angular velocity is  
	\begin{equation} 
		\Omega_{\mathrm{K}} = \sqrt{\frac{GM}{r^{3}}} \label{eq:1},
	\end{equation} 
	where $ M $ is the central object mass, $ r $ is the cylindrical radius, and $ G $ is the gravitational constant. The disk maintains hydrostatic equilibrium in the vertical direction, therefore the gravity and the pressure are balanced with each other, and the hydrostatic equilibrium equation is 
	\begin{equation} 
		\frac{dp}{dz} = - \frac{GM}{r^{3}} \rho z \label{eq:2}, 
	\end{equation} 
	where $ p $ and $ \rho $ are the pressure and the density, with $ z $, the vertical coordinate \citep{1977PThPh..58.1191H}. To describe the changing of $ p $ and $ \rho $ in the $ z $ direction, we employ the polytropic relation $ p = K \rho^{1 + 1/N} $ ($ K $ and $ N $ are constants) in the vertical direction \citep{1977PThPh..58.1191H, 2008bhad.book.....K}, and we assume $ N = 2 $ \citep{1995ApJ...452..710N, 1997ApJ...476...49N}. Using $ p = \lambda p_{0} \: (0 < \lambda < 1) $ at $ z = H_{1} $ ($ H_{1} $ is the half thickness of Disk 1) as the boundary conditions, we can get the solutions of $ p $ and $ \rho $ from Equation (\ref{eq:2}) for Disk 1, respectively expressed as $ p_{1} (z) $ and $ \rho_{1} (z) $: 
	
	\begin{equation}
		p_{1} (z) = p_{0} \cdot [1 - (1 - \lambda^{\frac{1}{N+1}} ) \frac{z^{2}}{H_{1}^{2}}]^{N+1}
		 \qquad (0 < z < H_{1}) \label{eq:3}, 
	\end{equation}

	\begin{equation}
		\rho_{1} (z) = \rho_{0} \cdot [1 - (1 - \lambda^{\frac{1}{N+1}} ) \frac{z^{2}}{H_{1}^{2}}]^{N} \qquad (0 < z < H_{1}) \label{eq:4}, 
	\end{equation}
	where $ p_{0} $ and $ \rho_{0} $ are the pressure and density at $ z = 0 $, respectively. By integrating Equations (\ref{eq:3}) and (\ref{eq:4}), we can get the pressure integration, $ \Pi_{1} = 2 \int_{0}^{H_{1}} p_{1} (z) dz $, and the surface density, $ \Sigma_{1} = 2 \int_{0}^{H_{1}} \rho_{1} (z) dz $.
	
	The pressures in Disk 1 at $ z = 0 $ include gas and radiation pressures, expressed as
	\begin{equation}
		p_{0} = \frac{\rho_{0} k_{\mathrm{B}} T_{0}}{\mu m_{\mathrm{H}}} + \frac{1}{3} a T_{0}^{4} \label{eq:5}, 
	\end{equation}
	where $ T_{0} $ is the temperature at $ z = 0 $ \citep{2008bhad.book.....K}, $ \mu $, $ m_{\mathrm{H}} $, $ k_{\mathrm{B}} $ and $ a $ respectively represent the mean molecular weight ($ \mu = 0.5 $), the hydrogen atom mass, the Boltzmann constant, and the radiation constant. From Equations (\ref{eq:2}), (\ref{eq:3}) and (\ref{eq:4}), we can get the equation for $ H_{1} $,
	\begin{equation}
	\Omega_{\mathrm{K}}^{2} H_{1}^{2} = 2(N+1)(1 - \lambda^{\frac{1}{N+1}}) \frac{p_{0}}{\rho_{0}}
        \label{eq:6}, 
	\end{equation}
	where $ 2(N+1)(1 - \lambda^{\frac{1}{N+1}}) $ is the integration constant.
	
	The continuity equation is
	\begin{equation}
		\dot{M}_{1} = - 2 \pi r v_{1} \Sigma_{1} \label{eq:7}, 
	\end{equation} 
	where $ \dot{M}_{1} $ is the accretion rate of Disk 1, and $ v_{1} $ is the radial velocity of Disk 1 \citep[Chapter 3.2]{2008bhad.book.....K}. The angular momentum equation is
	\begin{equation}
		\dot{M}_{1}(r^{2} \Omega_{\mathrm{K}} - j) = -2 \pi r^{2} T_{r\phi, 1} \label{eq:8}, 
	\end{equation}
	where $ j $ is the specific angular momentum per unit mass swallowed by the black hole, and $ T_{r\phi, 1} = -(\alpha \Pi_{1} + I) $ is the vertically integrated stress exerted on Disk 1, where $ I $ is the interaction between the two disks caused by the magnetic fields in the $ r \phi $-plane, and $ \alpha $ is the viscosity parameter.
	
	The thermal equilibrium of Disk 1 is expressed as
	\begin{equation}
		Q_{\mathrm{vis, 1}}^{+} = Q_{\mathrm{rad, 1}}^{-} \label{eq:9}, 
	\end{equation}
	where $ Q_{\mathrm{vis, 1}}^{+} $ is the viscous heating rate per unit area, and $ Q_{\mathrm{rad, 1}}^{-} $ is the radiation cooling rate per unit area, respectively. They are expressed as
	\begin{equation}
		Q_{\mathrm{vis, 1}}^{+} = r T_{r \phi, 1} \frac{d\Omega_{\mathrm{K}}}{dr} \label{eq:10}, 
	\end{equation}
	
	\begin{equation}
		Q_{\mathrm{rad, 1}}^{-} = \frac{8 ac T_{0}^{4}}{3 \tau_{1}} \label{eq:11}, 
	\end{equation}
	where $ \tau_{1} = \int_{0}^{H_{1}} \kappa_{1} \rho_{1} (z) dz $, and $ \kappa_{1} = \kappa_{\mathrm{es, 1}} + \kappa_{\mathrm{ff, 1}} $ is the opacity of Disk 1. We employ $ \kappa_{\mathrm{es, 1}} = 0.4 \: \mathrm{cm^{2} g^{-1}} $ and $ \kappa_{\mathrm{ff, 1}} = 6.4 \times 10^{22} \overline{\rho}_{1} \overline{T}_{1}^{-3.5} \: \mathrm{cm^{2} g^{-1}} $ as the electron scattering and free-free absorption opacities respectively, where $ \overline{\rho}_{1} $ and $ \overline{T}_{1} $ are the average quantities in the $ z $ direction \citep[Chapter 7.2]{2008bhad.book.....K}. Disk 1 is dominated by the electron scattering opacity, thus $ \kappa_{1} \sim \kappa_{\mathrm{es, 1}} $.
	
	\subsection{Optically Thin Disk (Disk 2)}
	
	We employ an optically thin accretion flow for Disk 2. This disk is geometrically thick, cooled by radiation and advection. The accretion rate of Disk 2 is lower than Disk 1, and thus the proportion of Disk 2 is $ \dot{M}_{2} / \dot{M} \leq 0.5 $, where $ \dot{M} = \dot{M}_{1} + \dot{M}_{2} $ is the total accretion rate and $ \dot{M}_{2} $ is the accretion rate of Disk 2.
	
	Equations (\ref{eq:1}) and (\ref{eq:2}) are also employed for the angular velocity and the hydrostatic equilibrium in $ z $ direction, respectively. Integrating Equation (\ref{eq:2}) from $ z = H_{1} $ to $ z = H_{1} + H_{2} $, with the same polytropic relation as Disk 1, and $ p = 0 $ at $ z = H_{1} + H_{2} $ as the boundary conditions, we can get the solutions of $ p $ and $ \rho $ for Disk 2, respectively expressed as $ p_{2} (z) $ and $ \rho_{2} (z) $: 
	\begin{equation}
		p_{2}(z)
	    = p_{\mathrm{2, b}}\cdot[1 - \frac{z^{2} - H_{1}^{2}}{(H_{1} + H_{2})^{2} - H_{1}^{2}}]^{N + 1} \qquad (H_{1} < z < H_{1} + H_{2}),  \label{eq:12}
	\end{equation}
	
	\begin{equation}
		\rho_{2}(z)
	    = \rho_{\mathrm{2, b}} \cdot[1 - \frac{z^{2} - H_{1}^{2}}{(H_{1} + H_{2})^{2} - H_{1}^{2}}]^{N} \qquad (H_{1} < z < H_{1} + H_{2}),  \label{eq:13}
	\end{equation}
	where $ p_{\mathrm{2, b}} $ and $ \rho_{\mathrm{2, b}} $ are the pressure and density at the bottom of Disk 2, respectively. To ensure the continuity of pressure in $ z $ direction, 
	$ p_{\mathrm{2, b}} = \lambda p_{0} $. In the $ z $ direction, $ p_{\mathrm{2, b}} $ corresponds to the maximum pressure for Disk 2, and $ \rho_{\mathrm{2, b}} $ corresponds to the maximum density. 
	
	The pressures in Disk 2 include gas and magnetic pressures. Assuming that $ \beta_{\mathrm{m}} = p_{\mathrm{mag, 2}} / p_{\mathrm{gas, 2}} $, the pressure of Disk 2 is
	
	\begin{equation}
	    p_{\mathrm{2, b}} = p_{\mathrm{gas, 2b}} + p_{\mathrm{mag, 2b}} = (1 + \beta_{\mathrm{m}}) p_{\mathrm{gas, 2b}}
        \label{eq:14},
	\end{equation}
    where $ p_{\mathrm{gas, 2b}} = \rho_{\mathrm{2, b}} k_{\mathrm{B}} T_{\mathrm{2, b}} / \mu m_{\mathrm{H}} $, and we assume that $ \beta_{\mathrm{m}} $ is constant in $ z $ direction. According to the simulation results by \cite{2014ApJ...784..169J}, and the theoretical investigation for the magnetic pressure in the warm corona by \cite{2015A&A...580A..77R}, a corona dominated by magnetic pressure can form above a gas pressure dominated disk, and $ p_{\mathrm{mag}} / p_{\mathrm{gas}} $ can exceed to 100 in the corona. Therefore we employ the value for $ \beta_{\mathrm{m}} $ ranging from 0 to 100, while in Disk 1 we assume that the magnetic pressure is negligible. From Equations (\ref{eq:2}), (\ref{eq:12}) and (\ref{eq:13}), we can get the equation for $ H_{2} $,
	\begin{equation}
		\Omega_{\mathrm{K}}^{2} (H_{2}^{2} + 2 H_{1} H_{2}) = 2(N+1) \frac{p_{\mathrm{2, b}}}{\rho_{\mathrm{2, b}}}  \label{eq:15}, 
	\end{equation}
	where the integration factor is $ 2(N+1) $. The pressure integration can be estimated as $ \Pi_{2} = 2 \int_{H_{1}}^{H_{1} + H_{2}} \: p_{2} (z) \: dz $, and the surface density can be estimated as $ \Sigma_{2} = 2 \int_{H_{1}}^{H_{1} + H_{2}} \: \rho_{2} (z) \: dz $.
	The continuity equation has the same form as Equation (\ref{eq:7}), expressed as
	\begin{equation}
		\dot{M}_{2} = - 2 \pi r  v_{2} \Sigma_{2} \label{eq:16},
	\end{equation}
	where $ v_{2} $ is the radial velocity for Disk 2. And the angular momentum equation is
	\begin{equation}
		\dot{M}_{2}(r^{2} \Omega_{\mathrm{K}} - j) = -2 \pi r^{2} T_{r\phi, 2} \label{eq:17}, 
	\end{equation}
	where $ T_{r\phi, 2} = -(\alpha \Pi_{2} - I) $ is the vertically integrated stress exerted on Disk 2. 
	
	The thermal equilibrium of Disk 2 is
	\begin{equation}
		Q_{\mathrm{vis, 2}}^{+} = Q_{\mathrm{adv, 2}}^{-} + Q_{\mathrm{rad, 2}}^{-} \: \label{eq:18},
	\end{equation}
	where $ Q_{\mathrm{vis, 2}}^{+} $ is the viscous heating rate per unit area, $ Q_{\mathrm{adv, 2}}^{-} $ is the advection cooling rate and $ Q_{\mathrm{rad, 2}}^{-} $ is the bremsstrahlung radiation cooling rate, respectively expressed as
	
	\begin{equation}
		Q_{\mathrm{vis, 2}}^{+} = r T_{r\phi, 2} \frac{d\Omega_{\mathrm{K}}}{dr} \label{eq:19}, 
	\end{equation}
	
	\begin{equation}
		Q_{\mathrm{adv, 2}}^{-} = \frac{\dot{M}_{2}}{2 \pi r^{2}} \frac{\Pi_{2}}{\Sigma_{2}} \xi
	    \label{eq:20}, 
	\end{equation}
	
	\begin{equation}
		Q_{\mathrm{rad, 2}}^{-} = Q_{\mathrm{brem}}^{-} = 0.23 \times (1.24 \times 10^{21} \hspace{0.1em} \rho_{\mathrm{2, b}}^{2} \hspace{0.1em} H_{2} \hspace{0.1em} {T_{\mathrm{2, b}}}^{1/2}) \ \mathrm{erg} \ \mathrm{s}^{-1} \ \mathrm{cm}^{-2} \label{eq:21}, 
	\end{equation}
	where $ \xi $ is
	\begin{equation}
		\xi = -[(A + \frac{1}{2}) \frac{d\ln\Pi_{2}}{d\ln r} - (A + \frac{3}{2}) \frac{d\ln \Sigma_{2}}{d\ln r} - \frac{d\ln \Omega_{\mathrm{K}}}{d\ln r}]
		\label{eq:22}, 
	\end{equation}
    with $ A = 3(1 - \beta) + \beta / (\gamma - 1) $, $ \beta = p_{\mathrm{gas, 2}} / p_{2} $, and $ \gamma = 5/3 $ \citep[Chapter 7.2]{2008bhad.book.....K}.

	\section{Numerical results} \label{sec:3}
	
	We propose the accretion timescale of Disk 1 to explain the CL timescale, expressed as $ t_{\mathrm{acc, 1}} = -r/v_{1} $. We assume that the UV/optical photons are emitted at $ r = 50 \: r_{\mathrm{s}} $, where $ r_{\mathrm{s}} = 2GM/c^{2} $ is the Schwarzschild radius \citep[e.g.,][]{2021ApJ...916...61F, 2023ApJ...958..146W}. We can numerically solve Equations (\ref{eq:1}), (\ref{eq:5}) $ \sim $ (\ref{eq:9}) and (\ref{eq:14}) $ \sim $ (\ref{eq:18}) for $ t_{\mathrm{acc, 1}} $ by employing the input parameters shown in Table \ref{tab:1}. 
	
\begin{figure}[]
		\centering
		\includegraphics[width=0.6\textwidth]{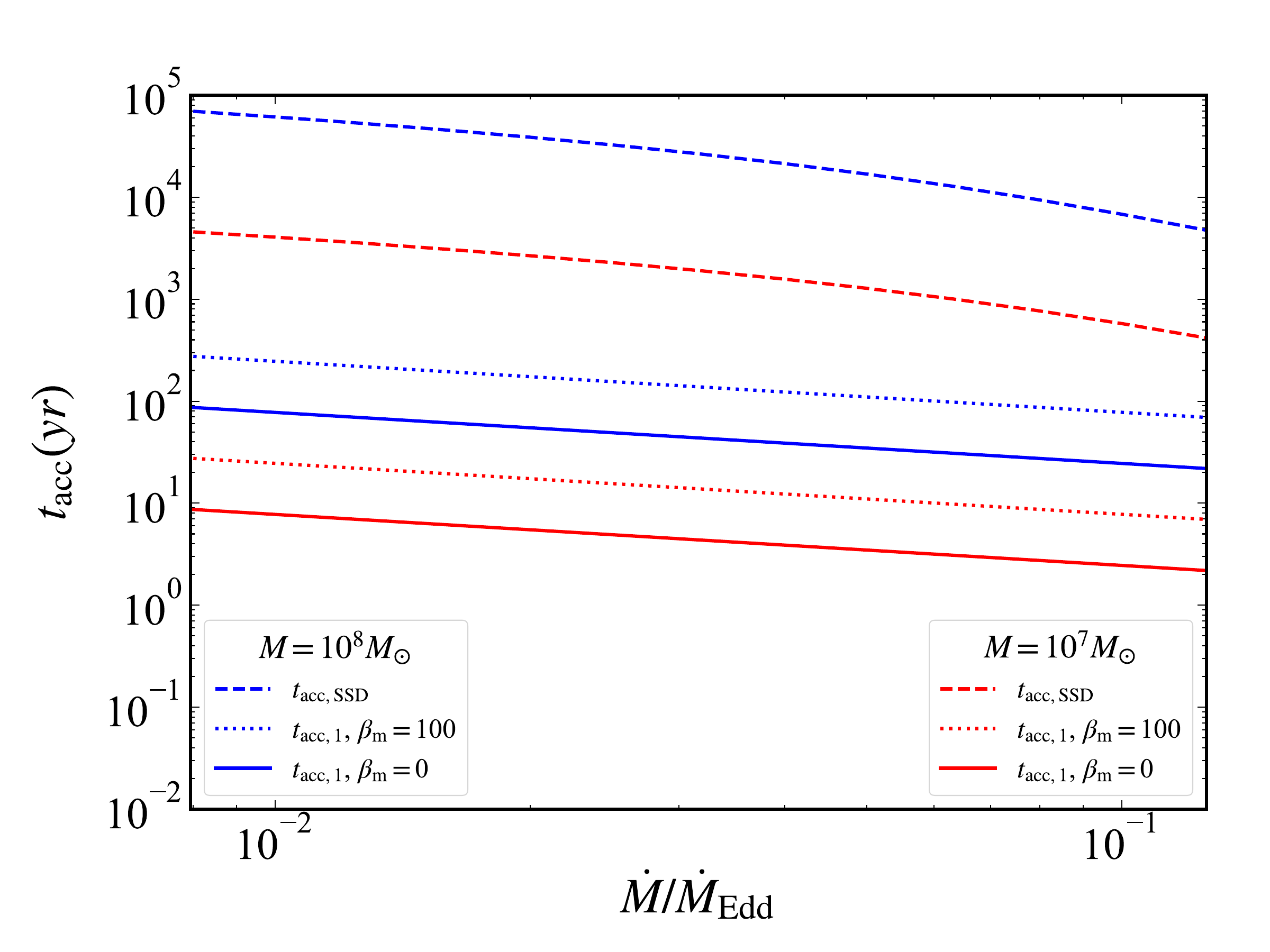}
		\caption{The accretion timescales for SSD and our Disk 1 at $ r = 50 \: r_{\mathrm{s}} $. In the timescale calculation, we fix $ \dot{M}_{2} / \dot{M} = 0.1 $ and $ v_{1} / v_{2} = 0.1 $. Compare to SSD, our Disk 1 has much shorter accretion timescales. }
		\label{fig:2}
	\end{figure}

	The comparison of the timescales between SSD and our model is shown in Figure \ref{fig:2}. For a typical AGN ($ M \sim 10^{7} - 10^{8} M_{\odot} $, $ \dot{M} \sim 0.01 - 0.1 \dot{M}_{\mathrm{Edd}} $) with a SSD, the accretion timescale is longer than $ 10^{3} $ yr, which is inconsistent with the observed CL timescale. In Figure \ref{fig:2}, we employ $ \dot{M}_{2} / \dot{M} = 0.1 $ and $ v_{1} / v_{2} = 0.1 $ to calculate $ t_{\mathrm{acc, 1}} $. The dotted and solid curves indicate that $ t_{\mathrm{acc, 1}} $ is two to three orders of magnitude shorter than SSD. For the condition that Disk~2 is dominated by the magnetic pressure ($ \beta_{\mathrm{m}} = 100 $), $ t_{\mathrm{acc, 1}} $ is relatively longer than the case of negligible magnetic pressure in Disk 2 ($ \beta_{\mathrm{m}} = 0 $). This result is contrary to \cite{2019MNRAS.483L..17D}, the magnetically elevated disk model, which shows that a magnetic pressure dominated disk will have a shorter accretion timescale. The reason for this lies in the difference of radiation mechanisms. The model proposed by \cite{2019MNRAS.483L..17D} is an optically thick disk, therefore it is primarily cooled by blackbody radiation, and its cooling rate is approximated as $ Q_{\mathrm{bb}}^{-} \propto T^{4} / \tau_{\mathrm{es}} \propto T^{4} / \Sigma $. Assuming that viscous dissipation is the only mechanism for angular momentum loss, the angular momentum equation is expressed as $ \dot{M} (r^{2} \Omega_{\mathrm{K}} - j) = 2 \pi r^{2} \alpha \Pi $, where $ \Pi = \Pi_{\mathrm{gas}} + \Pi_{\mathrm{mag}} + \Pi_{\mathrm{rad}} = constant $. Therefore an increasing $ \Pi_{\mathrm{mag}} $ will decrease $ \Pi_{\mathrm{gas}} $ and $ \Pi_{\mathrm{rad}} $, resulting to a lower temperature. Our Disk 2 is primarily cooled by the bremsstrahlung radiation, and employing $ \Sigma \propto \tau_{\mathrm{es}} \propto \rho H $, $ H_{2} \propto (1 + \beta_{\mathrm{m}})^{1/2} T_{2}^{1/2} $ (from Equations (\ref{eq:14}) and (\ref{eq:15}), in our model $ H_{2} \gg H_{1} $), the cooling rate is approximated as $ Q_{\mathrm{brem, 2}}^{-} \propto \rho_{2}^{2} H_{2} T_{2}^{1/2} \propto \Sigma_{2}^{2} / (1 + \beta_{\mathrm{m}})^{1/2} $. Here we assume that $ Q_{\mathrm{bb}}^{-} $ and $ Q_{\mathrm{brem}}^{-} $ are constant and are not affected by $ \beta_{\mathrm{m}} $. Therefore, with the increasing of magnetic pressure, for a disk cooled by $ Q_{\mathrm{bb}}^{-} \propto T^{4} / \Sigma $, an increase in $ \beta_{\mathrm{m}} $ will decrease $ T $, resulting to a decrease of $ \Sigma $, and considering that $ \Sigma \propto v_{r}^{-1} \propto t_{\mathrm{acc}} $, $ t_{\mathrm{acc}} $ will be shorter. However, for a disk cooled by $ Q_{\mathrm{brem}}^{-} \propto \Sigma^{2} / (1 + \beta_{\mathrm{m}})^{1/2} $, as $ \beta_{\mathrm{m}} $ increases, $ \Sigma $ also increases, which leads to a longer $ t_{\mathrm{acc}} $. Therefore, a magnetic pressure dominated Disk 2 will have a longer accretion timescale. As a result, under the condition of constant $ v_{1} / v_{2} $, the accretion timescale of Disk 1 also increases.

\begin{figure*}
		\gridline{\fig{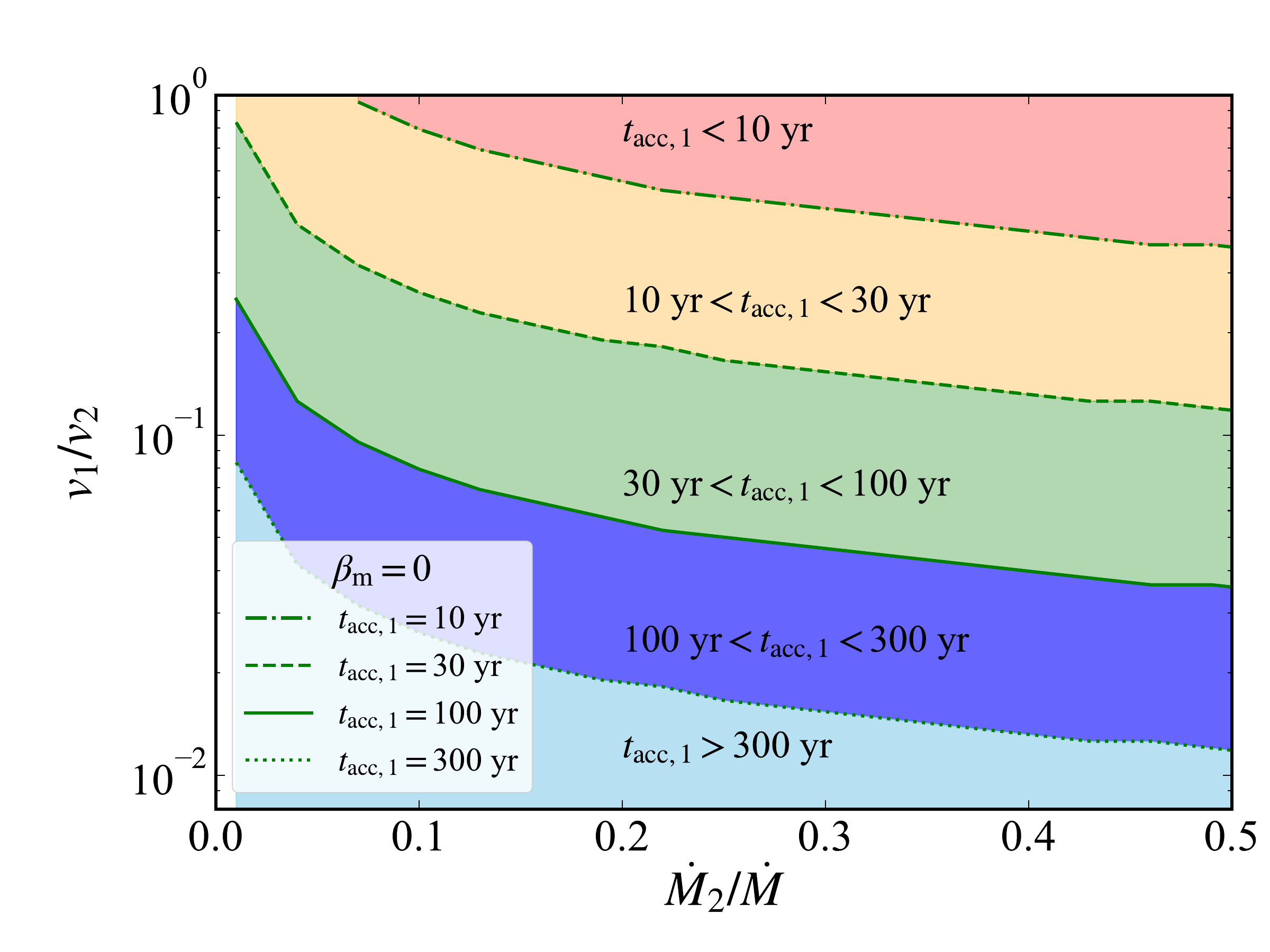}{0.45\textwidth}{(a)}
			\fig{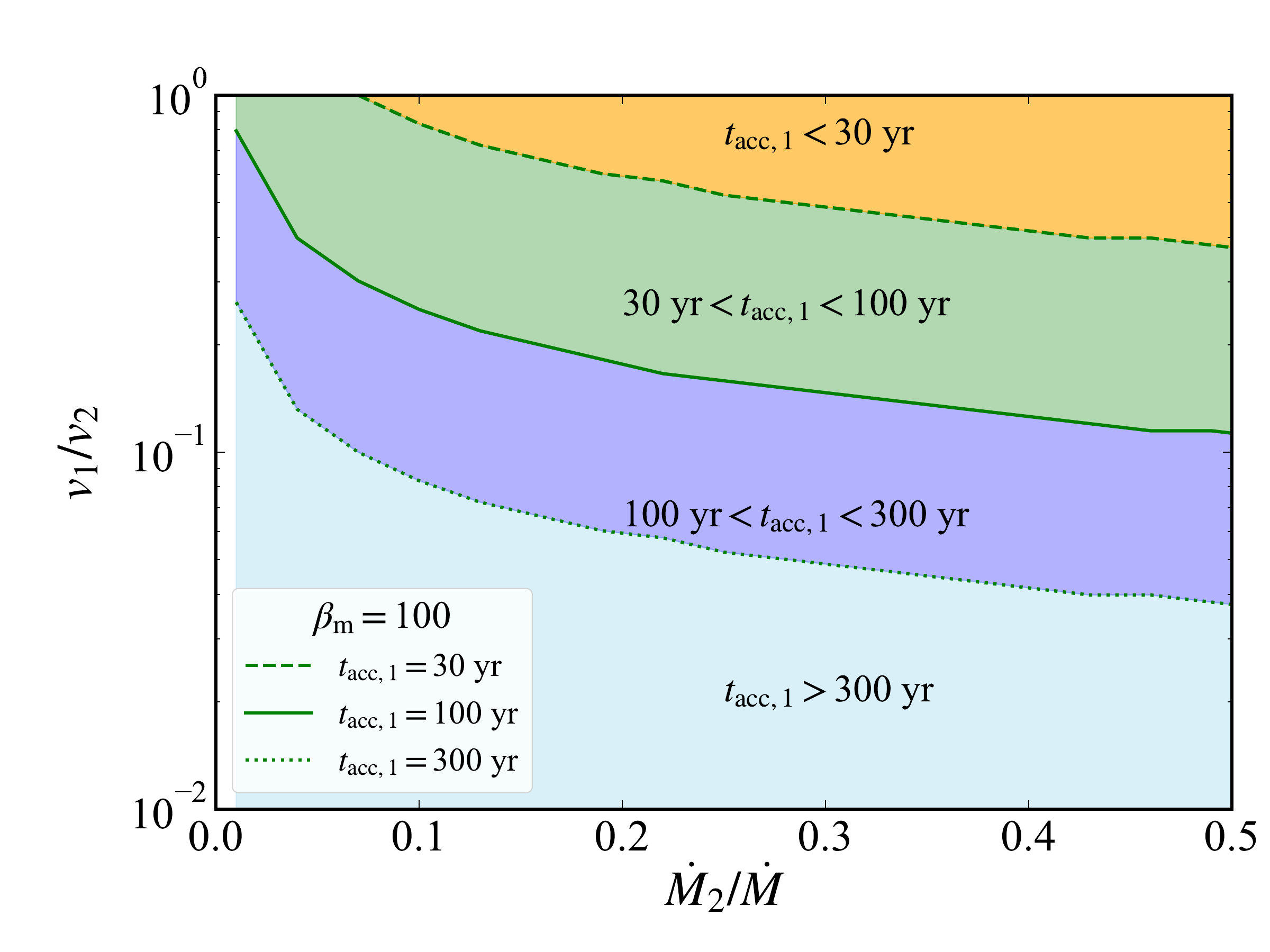}{0.45\textwidth}{(b)}
		}
		\caption{The ranges of $ t_{\mathrm{acc, 1}} $ as a function of $ v_{1} / v_{2} $ and $ \dot{M}_{2} / \dot{M} $. The other input parameters: $  M = 10^{8} M_{\odot} $, $ \dot{M} = 0.01 \dot{M}_{\mathrm{Edd}} $. The left and right panels correspond to gas pressure dominated and magnetic pressure dominated conditions, respectively.}
		\label{fig:3}
	\end{figure*}

	Figure \ref{fig:3} is the parameter space of $ v_{1} / v_{2} $ and $ \dot{M}_{2} / \dot{M} $ for given ranges of $ t_{\mathrm{acc, 1}} $. We adapt $ \dot{M} = 0.01 \dot{M}_{\mathrm{Edd}} $, $ M = 10^{8} M_{\odot} $ as the typical input parameters for this figure. The regions corresponding to different accretion timescales are represented by different colors. Figure \ref{fig:3}(a) is for the gas pressure dominated condition ($ \beta_{\mathrm{m}} = 0 $) while (b) is for the magnetic pressure dominated condition ($ \beta_{\mathrm{m}} = 100 $). The observed CL timescales can be easily explained by our model if Disk 2 is responsible for transporting the angular momentum of Disk 1, especially if Disk 2 is not magnetic pressure dominated. 
	
\begin{figure*}
		\gridline{\fig{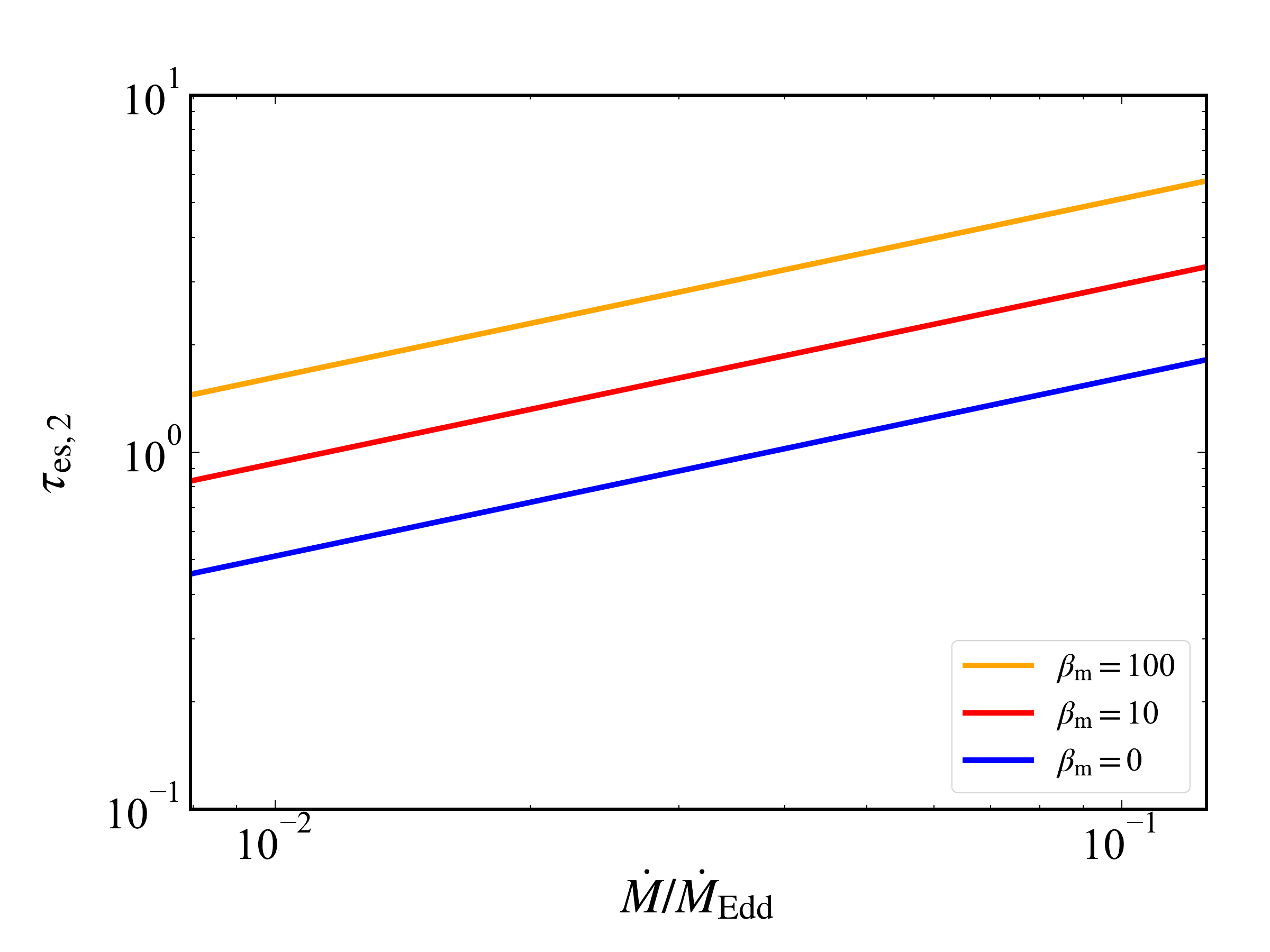}{0.45\textwidth}{(a)}
			\fig{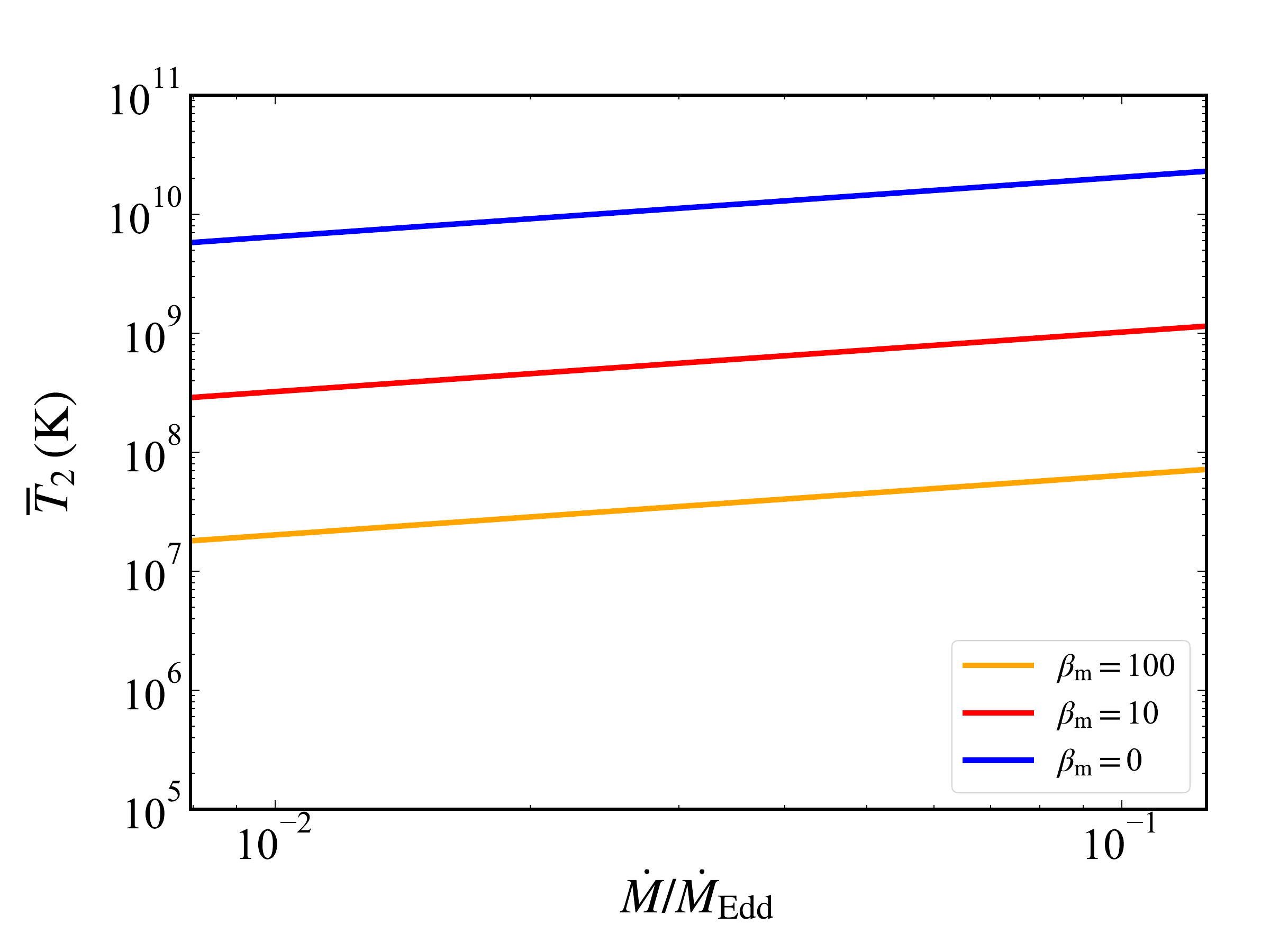}{0.45\textwidth}{(b)}
		}
		\caption{The electron scattering optical depth and average temperature of Disk 2 with different $ \beta_{\mathrm{m}} $. The other input parameters: $ M = 10^{8} M_{\odot} $, $ \dot{M}_{2} / \dot{M} = 0.1 $, and $ v_{1} / v_{2} = 0.1 $. }
		\label{fig:4}
	\end{figure*}

	Figure \ref{fig:4} shows the electron scattering optical depth and average temperature for Disk 2 with different $ \beta_{\mathrm{m}} $. Compared to the gas pressure dominated condition, the magnetic pressure dominated condition corresponds to greater $ \tau_{\mathrm{es, 2}} $ and lower $ \overline{T}_{2} $. \cite{2024MNRAS.530.1603B} systematically analyzed the X-ray spectra of 14 Type 1 AGNs and determined the temperature and electron scattering optical depth of the warm coronae in these sources, which are $ T\sim 10^7 $ K and $ \tau_{\mathrm{es}} > 1 $, respectively. Hence, our Disk 2 may account for the warm corona if Disk 2 is dominated by magnetic pressure.
	
	\vspace{10pt}

	\vspace{10pt}
	
	\vspace{10pt}

	\section{Conclusions and Discussion}  \label{sec:4}

	We have investigated a sandwich model to explain the timescales for CL-AGNs, including an optically thick disk (Disk 1) and two optically thin disks (Disk 2). In our model, Disk 1 is on the equatorial plane, with two optically thin disks above it on both sides. The interaction between Disk 1 and Disk 2 is generated by magnetic coupling, which causes Disk 1 losing angular momentum more quickly and decreases its accretion timescale. Our results indicate that, the accretion timescale for Disk 1 can be reduced by two to three orders of magnitude compared to SSD, which is responsible for the rapid variability of CL-AGNs. In addition, we also discuss the impact of magnetic pressure in Disk 2. When the magnetic pressure is negligible, Disk 2 manifests itself a ``hot corona" and has a shorter accretion timescale. And when the magnetic pressure is two orders of magnitude stronger than the gas pressure, Disk 2 resembles a ``warm corona'' and has a longer accretion timescale.
	
    The self-consistency of our model is verified by the effective optical depth $ \tau_{\ast} $. We have confirmed that within the parameter range shown in Table \ref{tab:1}, $ \tau_{\ast, 1} > 1 $ and $ \tau_{\ast, 2} < 1 $ can be ensured. For Disk 1, the cooling is dominated by the blackbody radiation for $ \tau_{\ast} > 1 $	\citep[e.g.,][]{1991ApJ...380...84W, 1995ApJ...452..710N}, while for Disk 2, $ \tau_{\ast} < 1 $ and the bremsstrahlung radiation is the main cooling mechanism. In addition, for $ t_{\mathrm{acc, 1}} < 100 $ yr, the stress $ I $ owing to the strong magnetic coupling is three to four orders of magnitude stronger than
	the viscous stress generated by Disk~1 ($I \gg \alpha \Pi_{1}$). We speculate that the magnetic coupling in most AGNs is not strong, and only in CL-AGNs, this mechanism is significant, which may explain the small proportion of CL-AGNs.
	
    In our model, the heating mechanism of Disk 2 is the viscous heating, which can be self-consistently calculated. Additionally, our model explains both the short timescale and the warm corona observed in CL-AGNs. The warm corona is associated with soft X-ray excess, which is usually observed in Type 1 AGNs \citep[e.g.,][]{2007MNRAS.374..150S, 2018MNRAS.480.3898N, 2022ApJ...937...31G, 2024MNRAS.530.1603B}, and during the transition from Type 2 to Type 1 the soft X-ray excess appears \citep[e.g.,][]{2018ApJ...866..123M, 2021MNRAS.507..687J, 2023A&A...670A.103K}. According to our model, we speculate that Disk 2 resembles the warm corona because of the increase of magnetic field and is responsible for the soft X-ray excess.
    \\

	We thank the anonymous referee for constructive suggestions that improved the paper. This work was supported by the National Natural Science Foundation of China under grants 12033006, 12433007, 11925301, 12221003, 12233007, and 12322303.

	
	
\end{document}